\documentstyle[aps,preprint,floats,epsf]{revtex}
\tighten
\draft
\begin{document}
\title{
Matter-Enhanced Neutrino Oscillations in the Quasi-Adiabatic Limit}
\author{A.~B.~Balantekin$^{1,2,3}$\thanks{Electronic address:
        {\tt baha@nucth.physics.wisc.edu}}, 
J.~F.~Beacom$^1$\thanks{Electronic address:
        {\tt beacom@citnp.caltech.edu}}
        \thanks{Present address: Physics 161-33, Caltech, Pasadena, CA
          91125 USA}
and J.~M.~Fetter$^1$\thanks{Electronic address:
        {\tt fetter@nucth.physics.wisc.edu}}}
\address{$^1$Department of Physics, University of Wisconsin\\
         Madison, Wisconsin 53706 USA\\
        $^2$Institute for Nuclear Theory, University of Washington, Box
         351550\\
         Seattle, WA 98195-1550 USA\\
         $^3$Department of Astronomy, University of Washington, 
          Box 351580\\
       Seattle, WA 98195-1580 USA}
\date{\today}
\maketitle

\begin{abstract}

We introduce simple analytic expressions for the neutrino survival
probability in media valid in the quasi-adiabatic limit. These
expressions provide a quick but accurate alternative to numerical
solution of the neutrino propagation equations for the
Mikheyev-Smirnov-Wolfenstein effect. They can also be used to extract
information about the density scale height from the neutrino data.
As an example, we present calculations for solar neutrinos.
\end{abstract}



\newpage

Neutrino flavor conversion via the Mikheyev-Smirnov-Wolfenstein (MSW)
mechanism \cite{msw} has been studied in various astrophysical
environments, most extensively in the sun \cite{parke,sun} and
supernov\ae \cite{SN}. It is straightforward to solve the appropriate
equations of motion analytically in some special cases \cite{baha2},
and numerically for an arbitrary density profile.  However, as one is
treating a quantum-mixing problem over considerable distances,
calculations which search a broad range of parameter space may quickly
become more tedious than this fundamentally simple level-crossing
problem should warrant. Furthermore, analytic approximations are
usually sufficient to highlight the salient physics.

A convenient starting point for most approximations is the formula
\cite{parke} for the averaged survival probability of a given neutrino
flavor: 
\begin{equation}
P_\nu(E) = 
\frac{1}{2}
\left[1 + (1 - 2 P_{hop})\cos{2\theta_v} \cos{2\theta_i}\right]\,.
\label{1}
\end{equation}
In this equation the vacuum- and matter-mixing angles are introduced
in the usual way, two-flavor mixing is assumed, and $P_{hop}$ denotes
the probability of hopping from one adiabatic eigenstate to the other.
One typically averages over the initial matter angle 
$\cos{2\theta_i}$ over the region of neutrino
production.  An excellent approximation for the hopping probability is
the quasi-adiabatic expression
\begin{eqnarray}
P_{hop} &=& \exp (- \pi \Omega ), \nonumber \\
\Omega &=& \frac{i}{\pi} \frac{\delta m^2}{2 E}
\int^{r_0^*}_{r_0} dr
\left[\zeta^2(r) - 2\zeta(r)\cos{2\theta_v} + 1\right]^{1/2}\,,
\label{2}
\end{eqnarray}
where ${r_0^*}$ and ${r_0}$ are the turning points (zeros) of the
integrand. In this expression we have defined 
\begin{equation}
\zeta(r) = \frac{2\sqrt{2} G_F N_e(r)}{\delta m^2/E}\,,
\label{3}
\end{equation}
where $N_e$ is the number density of electrons in the medium.  By
analytic continuation, this complex integral is primarily sensitive to
densities near the resonance point.  $P_{hop}$ varies
slightly over the production region, since neutrinos produced at
different places see slightly different variations of density along
their paths.  When the full three-dimensional geometry of the
production region is taken into account, this is a small but nonzero
effect.  For purposes of demonstration, however, we will neglect the
variation of $P_{hop}$ with production point.

Eq.~(\ref{2}) for $P_{hop}$ is valid for an arbitrary density
profile, and for a large range of mixing parameters.  In
Ref.~\cite{baha1}, a uniform semiclassical solution of the MSW
equations was introduced, and solved numerically.  In
Ref.~\cite{baha3}, a complete analytic solution for the oscillation
probability (including interference terms) was given, and the
expression for $P_{hop}$ above derived as a consequence.  Using a
different technique, the form of $P_{hop}$ above was found for the
exponential density in Ref.~\cite{piz}, and for the general case in
Ref.~\cite{pan}.  It is an excellent approximation from the adiabatic
regime up to the extreme non-adiabatic limit\cite{baha3}.  In
particular, it has a larger range of validity than the linear
Landau-Zener result, which can be recovered as a special case of
Eq. (\ref{2}). 

For the small-angle solution to the solar neutrino anomaly, the
propagation is nonadiabatic in the energy range of interest.  
$P_{hop}$ is appreciable (but not maximal) and is thus 
well-approximated by
Eq.~\ref{2}.  For this solution, $\cos{2\theta_i} = -1$ to a very good
approximation, so we have a complete analytic solution for $P_\nu(E)$.
The resonance condition
\begin{equation}
\frac{\delta m^2}{E} \cos{2\theta_v} = 2\sqrt{2} G_F N_e(r_{\rm res})
\label{res}
\end{equation}
relates the density probed by the resonance to a given neutrino
energy.  In Figure 1(a), we show the density profile from the 1995
standard solar model (SSM) of Bahcall and Pinsonneault\cite{ssm}. 
For the small angle solution, the
resonance for a 5 MeV neutrino occurs at about $0.35 R_\odot$; for
a 15 MeV neutrino it occurs at at about $0.45 R_\odot$.  In that
region, the solar density is approximately exponential.
This form of the density motivates an expansion of the electron number
density scale height, $r_s$, in powers of density:
\begin{equation}
 -r_s \equiv \frac{N_e(r)}{N_e'(r)} = \sum_n b_n N_e^n, 
 \label{4}
\end{equation}
where prime denotes derivative with respect to $r$. In this expression
a minus sign is introduced because we assumed that density profile
decreases as $r$ increases. For an exponential
density profile only the $n=0$ term is present. The $n \neq 0$ terms
represent deviations from the exponential profile.  The density scale
height is more sensitive to the fine structure of the medium than the
density profile itself.  Even for a density profile which appears
nearly exponential, the scale height may vary significantly from a
constant value.  To illustrate this we plot in Figure 1(b) the density
scale height for the standard solar model. 
In principle, for most monotonic density
profiles the scale height can be fitted using an expansion as given in
Eq.~(\ref{4}).  One should emphasize that, for an arbitrary density
profile, the expansion in Eq.~(\ref{4}) is not a perturbative
expansion.  That is, one should not expect the coefficients $b_n$ to 
necessarily decrease
as $n$ increases. Fits truncated in a given order could result in
coefficients which are rather different from coefficients obtained
with a different order, much like fitting a function with different
order splines.

Eq.~(\ref{4}) can easily be integrated to yield
\begin{equation}
  \label{4a}
  r(N_e) = r_b + b_0 \log N_e + \sum_{n \neq 0} \frac{b_n}{n} N_e^n\,,
\end{equation}
where $r_b$ is an integration constant, i.e., a set of $b_n$'s
specifies $N_e(r)$ up to a radial shift. Hence if only one of the
$b_n$'s in this expansion were non-zero, it would lead to $N_e(r) \sim
r^{1/n}$ for $n > 0$, and $N_e(r) \sim (1/r)^{1/|n|}$ for $n < 0$. In
principle, such $n < 0$ terms could appear, but they seem to be more
singular than necessary for modeling the solar density, hence in the
rest of this paper we take $n \ge 0$.\footnote{But they may be
necessary in other cases, e.g., for the density profile in a
core-collapse supernova, where this method is equally applicable.} In
practice one would fit a given density profile using a finite number
of terms, as illustrated in Figure 2.  The fit is only to be made over
the range of densities probed via Eq.~\ref{res} and the given range of
energy.  In this figure we show the density scale height of the
standard solar model in the region probed by the Sudbury Neutrino
Observatory (SNO)\cite{sno} and SuperKamiokande (SK)\cite{sk}
detectors.  We then fit it by truncating the expansion in
Eq.~(\ref{4}) at the $n=0,1,$ and $2$ term, respectively.  Just a few
terms of the series approximate the model's density profile quite
well.

Inserting the expansion of Eq.~(\ref{4}) into Eq.~(\ref{2}), and using an
integral representation of the Legendre functions \cite{ww}, one obtains
\begin{eqnarray}
\label{5}
\Omega &=& -\frac{\delta m^2}{2 E} \left\{ b_0 (1 - \cos{2\theta_v})
\frac{}{} \right. \nonumber \\  &+&  \left.  \sum^{\infty}_{n=1}
\left( \frac{ \delta m^2}{2 \sqrt{2} G_F E} \right)^n  \frac{b_n}{2n +
1}  \left[P_{n-1}(\cos{2\theta_v}) -
P_{n+1}(\cos{2\theta_v})\right]\right\}\,, 
\end{eqnarray}
where $P_n$ is the Legendre polynomial of order n.  The first term in
Eq.~(\ref{5}) represents the contribution of the exponential density
profile alone (as was implicit in the treatment in Ref.~\cite{Kwong}).
Eq.~(\ref{5}) is remarkable in that it directly connects an
expansion of the logarithm of the hopping probability in powers of
$1/E$ to an expansion of the density scale height.  That is, after a
slight amount of computation, it provides a direct connection between
$N_e(r)$ and $P_\nu(E_\nu)$.  Consequently these equations provide a
quick and accurate alternative to numerical integration of the MSW
equation for any density profile, for a wide range of mixing
parameters.  To illustrate the utility of Eq.~(\ref{5}) in Figure 3,
we compare the electron neutrino survival probability in the Sun
obtained using this equation with the exact numerical calculation.  To
generate the approximate survival probabilities in this figure, we
used the fit coefficients from Figure 2 and computed $\cos 2\theta_i$
numerically.   Eq.~(\ref{5}) is an
excellent approximation; the relative error in $P_\nu$ is well under
1\% for the order 2 fit, across the full range of neutrino energy
probed by SNO and SK.  We fit the $r_s$ profile only for densities
where neutrinos in our energy range undergo resonance. 

The SNO and SK detectors are sensitive to the spectrum distortion due
to the MSW effect. In Figure 4, we calculate
the spectrum distortion for the small angle MSW solution at SNO, using
the method of Ref.~\cite{snospec} and neglecting backgrounds.  For the
calculation of the neutrino-deuterium charged-current cross-sections
we used the code of Bahcall and Lisi \cite{bl}. This encoded
differential charged-current cross-section is based on the
Ellis-Bahcall effective range calculation \cite{eb} with slight
improvements.  One observes that even the first term in
Eq.~(\ref{5})---that is, the pure exponential---is sufficient to
describe five years of data collection, and to show the distortion
from the no-oscillation spectrum.  The success of using a single term 
depends, however, on the use of the correct scale height in the range
of resonance densities probed by MSW.  This scale height may be quite
different from the scale height that fits the entire sun with an
exponential $N_e(r)$.  For our mixing parameters, the fit to the full
sun would give a significantly different result, as noted below.

The MSW effect is an energy- and density-dependent effect.  With the
analytic forms given above, this is explicit for an arbitrary density
profile.  From the measured energy dependence of neutrino data, one
can use these expressions to invert for the scale height as a function
of density and hence density as a function of radius (up to a radial
shift).  
Data from SNO, in principle, would be particularly suitable for
such an analysis for two reasons. First, the electron spectra is the
charged-current is sharply peaked at the neutrino energy. Second, it
may be possible to use the neutral-current measurement to determine 
total $^8$B solar neutrino flux.  
The range of densities probed depends on the
range of energies probed, via Eq.~\ref{res}.  As noted above, an
energy range of $E = 5 - 15$ MeV probes a radial range of
approximately $0.35 R_\odot - 0.45 R_\odot$ if the small-angle
solution to the solar neutrino anomaly is assumed.  These resonance
positions are far outside the region where the high-energy $^8$B 
neutrinos are produced (peak production is at approximately $0.05
R_\odot$).  To a large extent, the finite size of the source can be
ignored, and only the energy distribution of these neutrinos is
needed.  This is not exact, as neutrinos created in a finite spherical
source
region and traveling to the earth will see slightly different
variations of density along their paths, and will hence have slightly
different hopping probabilities.  The primary systematic limitations
of this inversion are then the accuracy of the analytic form for
$P_{hop}$ and the treatment of a finite source region. 
  
However, the strongest limitation to making such an inversion is the
statistical error.  Thus probing the solar interior via the MSW effect
will require a sensitivity that goes beyond that of current solar
neutrino detectors, as we discuss below.  In Figure 5, we show what
SNO may accomplish with regard to inversion  in five years of running
time. 
To obtain the confidence regions in
this figure, we predicted the electron energy spectrum from
charged-current deuteron breakup at SNO after 5 years of counting. We
found the electron neutrino survival probabilities by full-fledged
numerical integration of the MSW equation through the standard solar
model of Bahcall and Pinsonneault.  Again, we used the method of
Ref.~\cite{snospec}, which takes account of SNO's finite electron
energy resolution.  We then randomly generated 2000 electron energy
spectra, Poisson-distributed about the theoretical expectation. We fit
each of the 2000 statistical instances with Eq.~(\ref{5}), and used
that the fit coefficients to produce 2000 measured scale height
profiles.  Then the CL\% confidence region in the figure encloses the
measured profiles for the CL\% of sample runs whose spectra, before
fitting, have the smallest $\chi^2$ with respect to the original
spectrum.

Figure 5a shows the results of this procedure when we truncated the
series in Eq.~(\ref{5}) after $n=0$, that is, we treated the electron
density profile as a pure exponential.  In Figure 5b we took account
of first-order deviations from exponential, truncating the series
after $n=1$.  In both figures, the solid line shows the scale height
as a function of density in the SSM, while the long-dashed line in the
center shows the result of fitting the original spectrum.  Where the
solid line goes outside the confidence regions in Figure 5a, it simply
means that we are not very sensitive to the scale height at those
densities.  Figure 5b clearly shows the region of sensitivity; we can
determine the scale height with some accuracy around $2 \times
10^{24}$ cm$^{-3}$, but not elsewhere.  This is where neutrinos of
energy near 10 MeV undergo resonance, and those neutrinos contribute
the most to the observed electron spectrum.

Quite different scale-height profiles can give rise to rather similar
survival probabilities. This makes inversion a particularly
challenging problem as the size of the confidence regions in the
figure illustrates.  Note, however, that we can
easily rule out the value of scale height which fits the entire sun as
an exponential, $6.6 \times 10^9$ cm.

In conclusion, we have presented simple analytic expressions for the
neutrino survival probability in media. These expressions where the
energy dependence of the hopping probability is related to the
density-scale height in the medium are valid in the quasi-adiabatic
limit. To demonstrate the technique for the solar neutrinos, we
neglected effects arising from the three-dimensional neutrino source
geometry. However, one may easily incorporate them by fitting the
density profile separately for each ray from the neutrino production
region to the earth. Since ordinarily one must numerically integrate
the MSW equations for each ray, this technique still vastly speed up
the computation of $P_{\nu}$.  As long as the density profile is
monotonic in the resonance region  
and is not subject to fluctuations such as those described
in Ref.~\cite{fluc}, Eq. (\ref{5}) provides a quick and accurate
alternative to numerical solution of the neutrino propagation
equations for the Mikheyev-Smirnov-Wolfenstein effect. Its validity is
not necessarily restricted to the Sun; they can also be used to
calculate the MSW effect for type II supernov\ae\ and the day-night
effect in Earth.  Used in reverse, these approximations in principle
allow one to invert neutrino data for the density profile.  However,
the statistics of the current and near-term neutrino detectors are not
adequate to do this in a precise way.


\section*{ACKNOWLEDGMENTS}

We thank members of the University of Washington SNO group 
for useful discussions and J.N. Bahcall and E.
Lisi for permission to use their computer code for the calculation of
the neutrino-deuterium cross sections.  This work was supported in
part by the U.S. National Science Foundation Grant No.\ PHY-9605140 at
the University of Wisconsin, and in part by the University of
Wisconsin Research Committee with funds granted by the Wisconsin
Alumni Research Foundation.  We thank the Department of
Energy's Institute for Nuclear Theory and Department of Astronomy at
the University of Washington for their hospitality and Department of
Energy for partial support during the completion of this work.  J.F.B.
thanks Caltech for support as a Sherman Fairchild Postdoctoral Scholar
in Physics during the final stages of this work.



\newpage
\centerline{\bf Figure Captions}

\bigskip
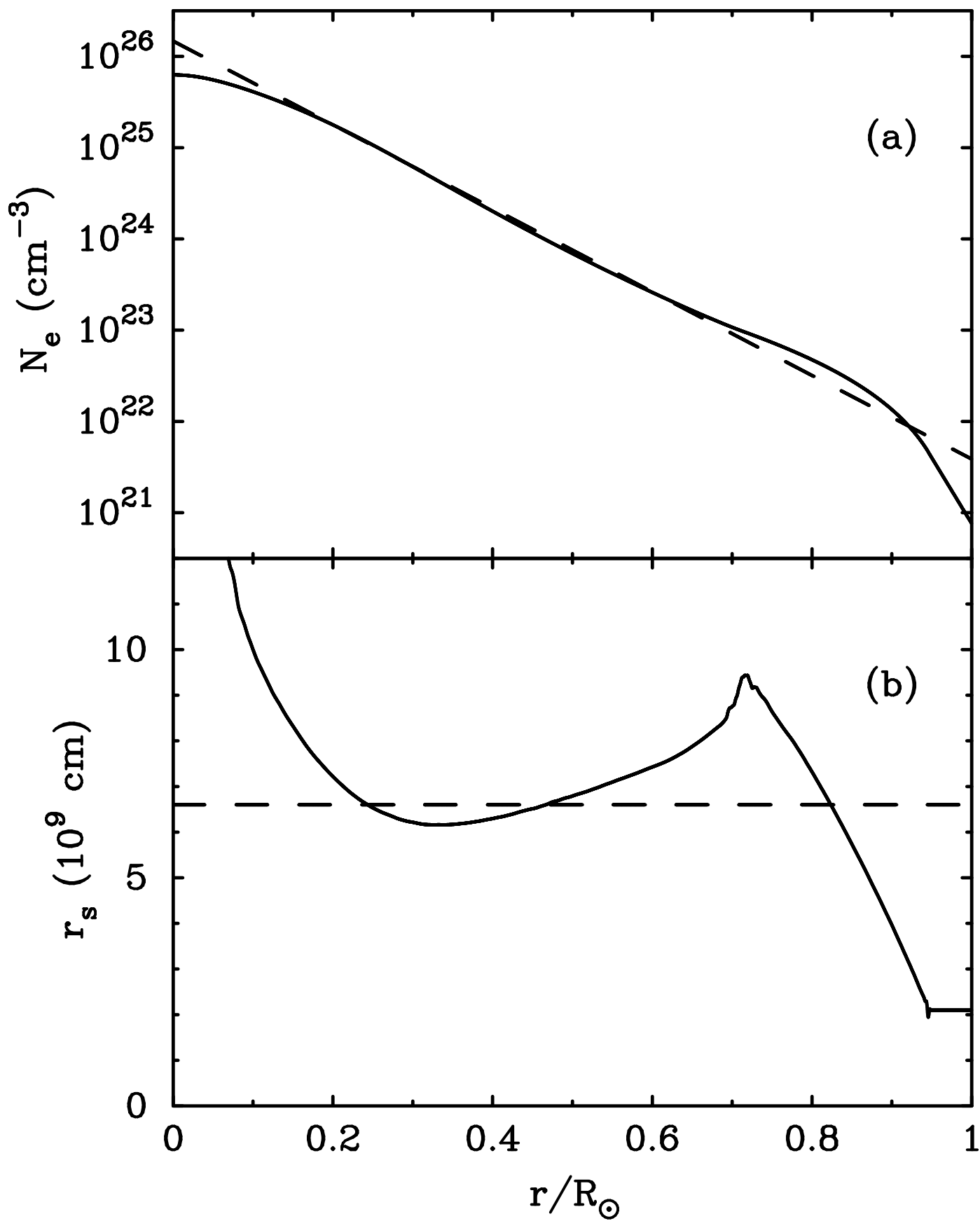
Figure 1.
(a) Density as a function of radius in the 1995 standard solar model
of Bahcall and Pinsonneault \cite{ssm}, shown with a solid line.  The
long-dashed line is the exponential fit over the whole sun.
(b) Scale height as a function of radius in the same model, shown with
a solid line.  The long-dashed line is the exponential fit over the
whole sun. The derivatives are obtained by smoothed splines from the
tables in Ref. \cite{ssm}. 

\bigskip 
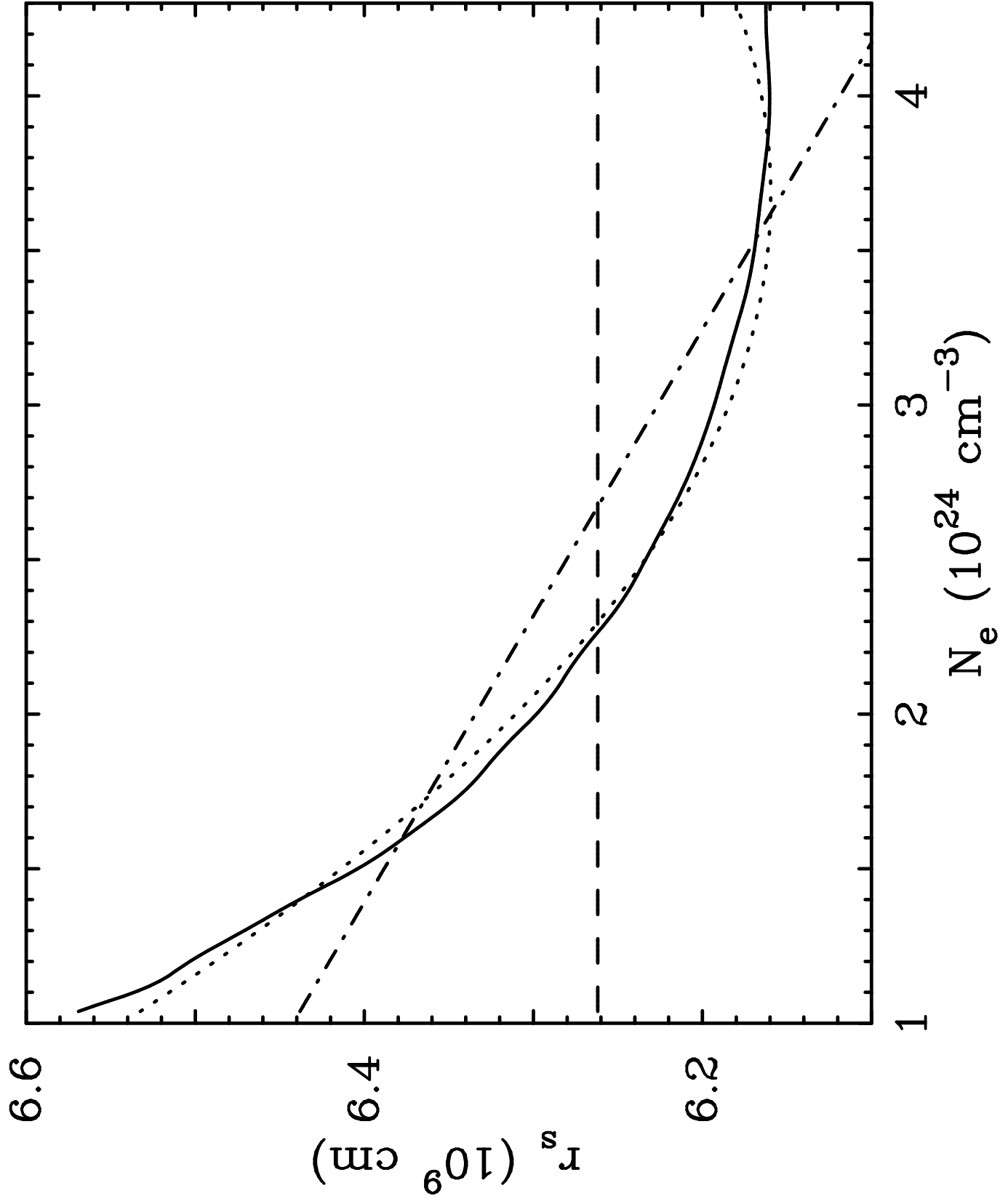
Figure 2. 
Scale height as a function of density in the standard solar model, in
the region probed by SNO and SK with the mixing parameters
$\sin{2\theta_v}= 0.01$ and $\delta m^2 = 5.0 \times 10^{-6}$ eV$^2$.
The solid line is the SSM value. The dashed, dot-dashed, and dotted
lines are obtained by using the expansion in Eq.~(\ref{4}) with values
of $n$ up to 0, 1, and 2, respectively.

\bigskip
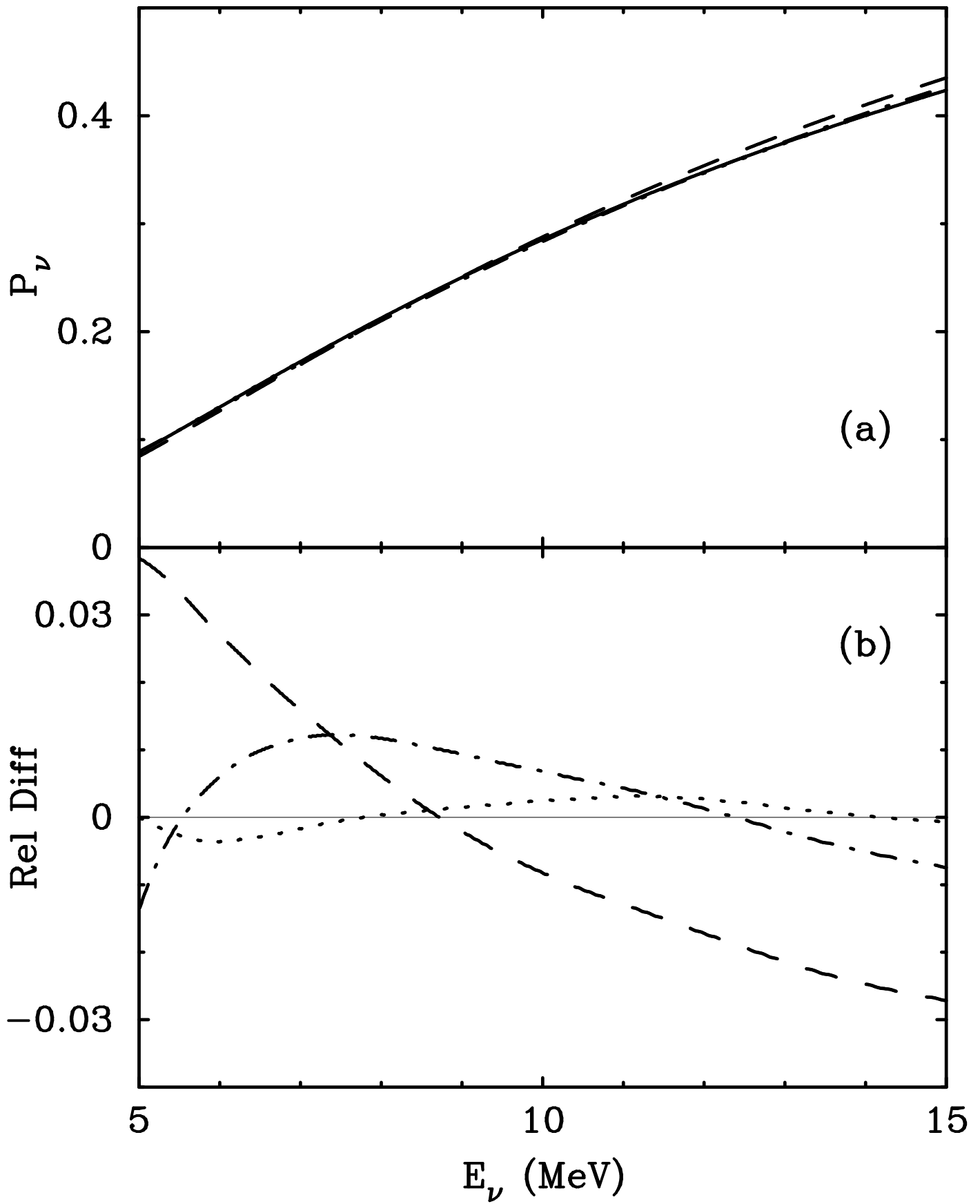
Figure 3.
(a) Electron neutrino survival probability in the SSM, as a
function of neutrino energy. The neutrino mixing parameters are the same as
in Figure 2.  The dashed, dot-dashed, and dotted lines are obtained from
Eq.~(\ref{5}) with values of $n$ up to 0, 1, and 2, respectively, and with
the parameters $b_n$ determined from the fits in Figure 2.
(b) The relative error, (exact - approx) / exact, introduced by using
Eq.~(\ref{5}) rather than a full numerical integration.

\bigskip 
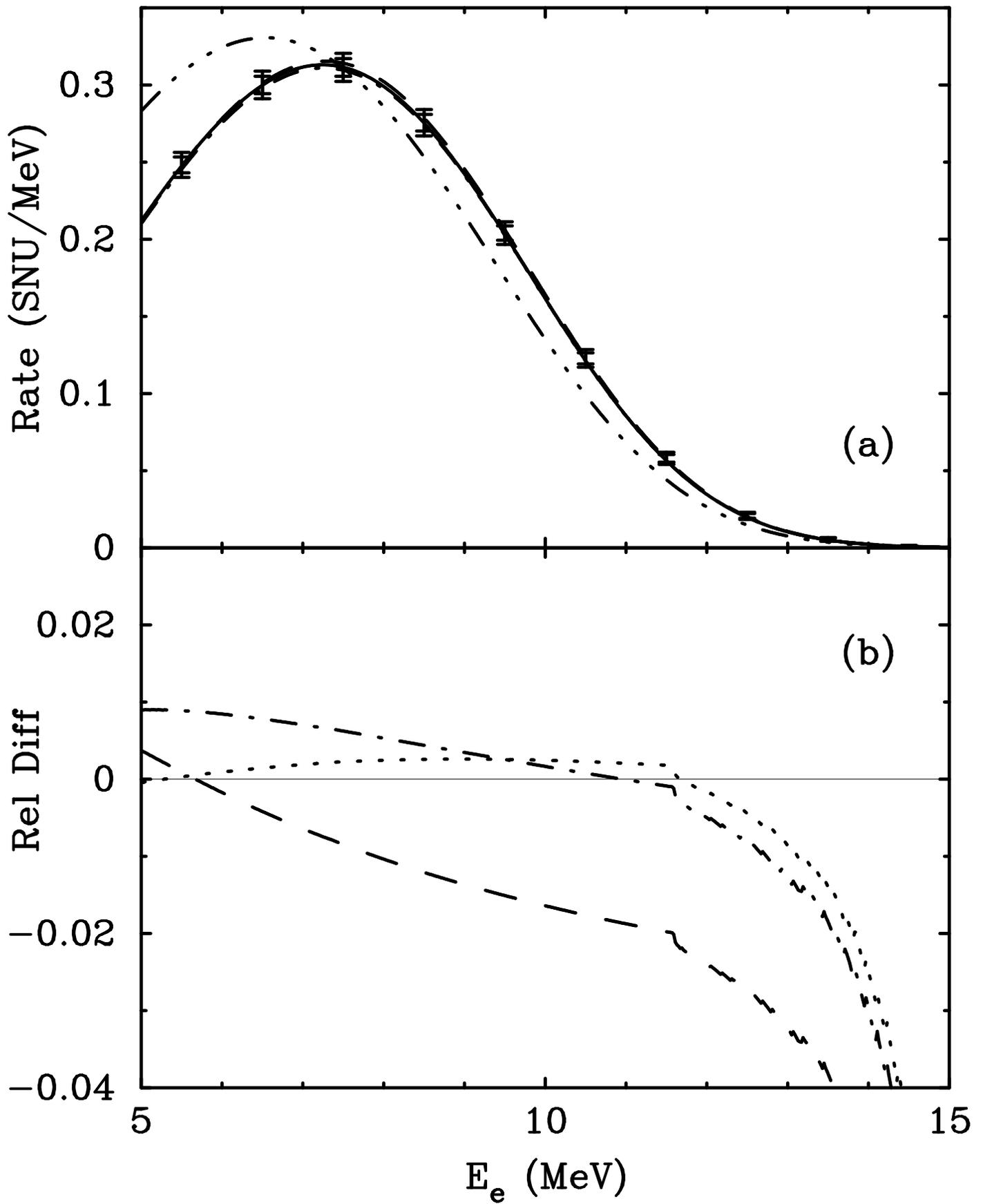
Figure 4.
(a) Spectrum distortion for the small-angle MSW solution at SNO,
resulting from the survival probabilities of Figure 3.  The solid line
is the exact numerical solution.  The dashed, dot-dashed, and dotted
lines result from values of $n$ up to 0, 1, and 2, as in Figure 3.
The error bars on the exact numerical result correspond to two and
five years of data collection.  The dot-dot-dot-dashed line is the
spectrum without MSW oscillations, normalized to the same total rate as
with MSW oscillations. Note that on the scale of this graph 
the $n=1$ and $2$ lines are not
distinguishable from the exact answer. 
(b) The relative error arising from the use of Eq.~(\ref{5}).

\bigskip
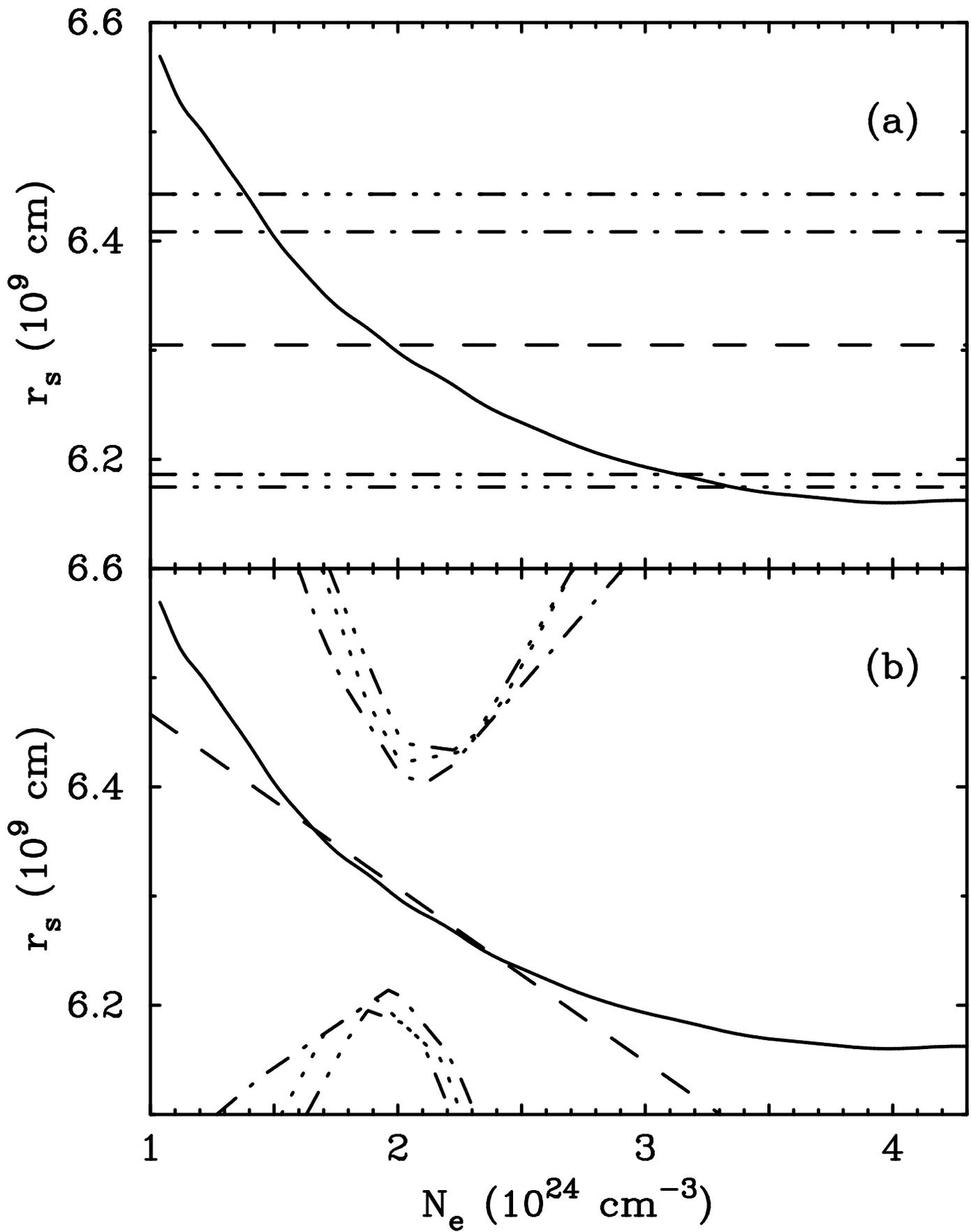
Figure 5.
(a) A determination of the density scale height in the range of
densities probed by MSW after five years of data collection at SNO.
The solid line is the scale height as a function of density in the
SSM.  The long-dashed line in the center of the figure is the scale
height which best fits the resulting electron energy spectrum, taking
only the $n=0$ term in Eq.~(\ref{5}).  The dot-dashed line shows the
boundary of the 68\% confidence region for scale height, and the
dot-dot-dot-dashed line shows the 99\% confidence region.  A dotted
line which shows the 90\% region is not visible in the figure; the
region's upper limit is the same as the 68\% region, and the lower
limit is the same as the 99\% region.
(b) As (a), but including the $n=1$ term in Eq.~(\ref{5}).  We now see
clearly the region of sensitivity in density, which corresponds to the
resonance densities for neutrinos that contribute significantly to the
observed electron spectrum.

\end{document}